\documentclass[%
 twocolumn,
superscriptaddress,
 amsmath,amssymb,
 aps,
]{revtex4-2}

\usepackage{graphicx}% Include figure files
\usepackage{dcolumn}% Align table columns on decimal point
\usepackage{bm}% bold math
\usepackage{float}
\usepackage{color}
\usepackage{textgreek}
\usepackage[separate-uncertainty=true,multi-part-units=single,list-units=repeat]{siunitx}
\usepackage[version=4]{mhchem}
\newcommand\blue[1]{{\color{blue}#1}}
\newcommand{\bld}[1]{\boldsymbol{#1}}
\usepackage{color}
\begin{document}

\preprint{APS/123-QED}

\title{ Quantifying nanoscale charge density features of contact-charged surfaces with an FEM/KPFM-hybrid approach }

\author{Felix Pertl}
  \email{felix.pertl@ist.ac.at}
\affiliation{Institute of Science and Technology Austria, Am Campus 1, 3400 Klosterneuburg, Austria}
\author{Juan Carlos Sobarzo}%
\affiliation{Institute of Science and Technology Austria, Am Campus 1, 3400 Klosterneuburg, Austria}
\author{Lubuna Shafeek}
\affiliation{Institute of Science and Technology Austria, Am Campus 1, 3400 Klosterneuburg, Austria}
\author{Tobias Cramer}
\affiliation{Department of Physics and Astronomy
University of Bologna,
Viale Berti Pichat 6/2, 40127 Bologna, Italy}
\author{Scott Waitukaitis}
\affiliation{Institute of Science and Technology Austria, Am Campus 1, 3400 Klosterneuburg, Austria}

\date{\today}

\begin{abstract}

Kelvin probe force microscopy (KPFM) is a powerful tool for studying contact electrification (CE) at the nanoscale, but converting KPFM voltage maps to charge density maps is non-trivial due to long-range forces and complex system geometry. Here we present a strategy using finite element method (FEM) simulations to determine the Green’s function of the KPFM probe/insulator/ground system, which allows us to quantitatively extract surface charge. Testing our approach with synthetic data, we find that accounting for the AFM tip, cone and cantilever are necessary to recover a known input, and that existing methods lead to gross miscalculation or even the incorrect sign of the underlying charge. Applying it to experimental data, we demonstrate its capacity to extract realistic surface charge densities and fine details from contact charged surfaces. Our method gives a straightforward recipe to convert qualitative KPFM voltage data into quantitative charge data over a range of experimental conditions, enabling quantitative CE at the nanoscale.

\end{abstract}

\maketitle

\section{\label{sec:intro} Introduction }

\blue{Contact electrification, the transfer of electric charge between objects during contact, is a ubiquitous and widely studied phenomenon, yet poorly understood \cite{Lacks.2019}. Surprisingly, the surface charge left on insulators after CE is heterogeneous, with correlated charge features spanning from nanometers to centimeters \cite{Shinbrot.2008, Burgo.2012, Moreira.2020ikk, Baytekin.2011, Sobolev.2022}. KPFM is the state-of-the art tool for addressing such features at the nanoscale, having revealed polarity-inverting `mosaics' \cite{Baytekin.2011}, diffusive surface dynamics \cite{Bai.2021} and correlations between charge polarity and mechanical deformation \cite{Li.2018}. Yet there is a catch---KPFM measures a voltage \textit{related} to surface charge, but not the charge itself, and conversion between the two is an unresolved issue. The challenge is twofold. First, although the necessary insights are peppered around the literature \cite{Orihuela.2016, Barth.2010, Gonzalez.2017, neff2015insights, Melin.2010}, there is no widespread understanding of what the physical relationship between surface charge and KPFM voltage is.  Second, even if that relationship is understood, one must account for long-range electrostatic forces that act over the many scales of complex AFM geometry, including spherical tips with radii on the order of \SI{10}{\nm}, conical probes with lengths on the order of \SI{10}{\um}, suspending cantilevers with dimensions of hundreds of microns, and sample thicknesses ranging from nanometers to millimeters. Many experiments, if not most, do not attempt to extract charge, and instead just report the KPFM voltage as a proxy \cite{Baytekin.2011, Bai.2021, Knorr.2011, ji2021stability}.  Unfortunately doing so leaves quantitative models for the origin of surface charge heterogeneity untestable \cite{Sobolev.2022}.  In some cases \cite{Barnes.2016, Wei:2016sr, knorr2012charge}, the charged surface and ground plane are thought of as a capacitor and the KPFM signal is presumed to be the voltage across this, but this intuitive approach has no rigorous physical backing, and cannot be expected at all to work for small charge features. 
\begin{figure}[b!]
\centering
\includegraphics[width=0.45\textwidth]{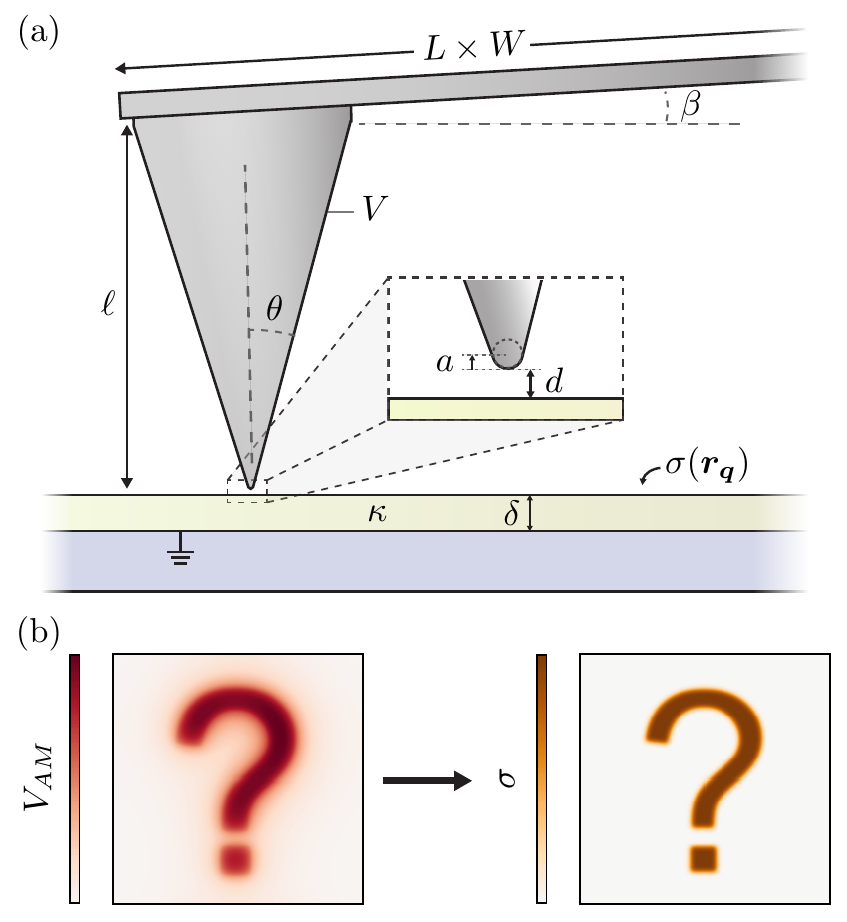}
\caption{ \label{fig:physical_setup} (a) An AFM probe of tip radius $a$, cone length $\ell$ and half angle $\theta$ is suspended by a cantilever of length $L$, width $W$, thickness $T$ and inclination $\beta$. The probe and cantilever are conductors at adjustable potential $V$. The tip is at distance $d$ above an insulator of thickness $\delta$ and relative permittivity $\kappa$. The bottom of the insulator is at ground, and the top has a varying surface charge density, $\sigma(\bld{r_q})$. (b) In KPFM, one measures a potential map, $V_{AM/FM}$, that is caused by the surface charge, but in a highly non-trivial way. Here, we present a strategy for the inverse problem, \textit{i.e.}~from a KPFM voltage map, extract the underlying charge density.} 
\end{figure}
Analytical approaches to convert voltage to charge have relied on aggressive simplifications, \textit{e.g.}~approximating the entire micron-scale AFM probe/cantilever by a nano-scale sphere \cite{Orihuela.2016, Gonzalez.2017}. In other cases, charge has been estimated with brute force numerics, \textit{e.g.}~discretizing the surface into point charges and adjusting their values until the KPFM voltage at one single location is reached \cite{Li.2018, ji2021mechano}. }

\blue{In this work, we present a method to convert KPFM voltage maps to surface charge density maps that is built on a rigorous physical basis and takes into account long-range electrostatic forces and the complex AFM geometry [Fig.~\ref{fig:physical_setup}(a,b)]. We focus on amplitude modulated (AM-)KPFM as it better suits our situation of interest, but in the Supplemental Material we explain how to implement it for frequency modulated (FM-)KPFM.  Guided by insights from Refs.~\cite{Orihuela.2016, Gonzalez.2017}, we first clarify that the key to the problem is to find the appropriate Green's function, which neatly casts the voltage map as a convolution over the surface charge. We show how this Green's function is related to different force terms of the system, and leverage this to construct it in FEM simulations. With the Green's function, converting from voltage to surface charge is a straightforward matter of deconvolution. We test our method with synthetic data and confirm it recovers the correct charge density when the input is known. These tests further reveal that existing methods can grossly miscalculate the charge magnitude, and in certain circumstances can even produce the incorrect sign of charge. Applying our method to experimental data, we demonstrate its capacity to extract realistic values for charge density from contact-charged surfaces.}

\section{\label{sec:theory} Theoretical basis }

The geometry we consider is illustrated in Fig.~\ref{fig:physical_setup}(a). An AFM probe with tip radius $a$, cone length $\ell$, and half angle $\theta$ is separated a height $d$ above an insulator layer of thickness $\delta$ and relative permittivity $\kappa$. The probe is held by a cantilever of length $L$, width $W$, thickness $T$, and inclination $\beta$.
The electrode below the insulator is at ground, while the probe and cantilever are at an adjustable potential $V$. \blue{Regarding charges in the system, we must make assumptions about what is present before/after CE and where. Before CE, we assume trapped charges may be present in the bulk and at the surface, but (1) they are homogeneously distributed and (2) the net charge is zero.  As we will show, we can ensure these conditions experimentally.  After and as a consequence of CE, we assume a thin layer of spatially varying surface charge, $\sigma(\bld{r_q})$, is present.  By `thin', we mean confined to a region near the surface whose thickness is small compared to the tip/sample distance, $d$. For materials similar to ours, this assumption is validated by the stability of charges on extremely thin substrates \cite{Chatelain.2021}, and by the observance of dominant lateral diffusion \cite{Bai.2021}.}

Guided by Refs.~\cite{Orihuela.2016, Barth.2010, Gonzalez.2017, neff2015insights}, we first consider the electrostatic energy of the system when a single point charge, $q$, is on the surface. Without loss of generality, this can be written as
\begin{equation}
    U = u_0q^2 + u_1qV + u_2V^2.
    \label{eq:energy}
\end{equation}
The first term, $u_0q^2$, comes from the charge interacting with its images in the cantilever/probe/insulator/ground capacitor (hence $\propto q^2$), and depends on the lateral distance from the tip, $|\bld{r_t}-\bld{r_q}|$, and geometric parameters, ${\cal G} = \{ \delta, \kappa, a, \theta, \ell, L, W, T, \beta \}$, \textit{i.e.}~$u_0 = u_0(|\bld{r_t}-\bld{r_q}|, {\cal G})$. The second term comes from the charge's interaction with the capacitor field ($\propto qV$), and thus it can be reasoned as $u_1=u_1(|\bld{r_t}-\bld{r_q}|, {\cal G})$. The final term is the energy of the capacitor itself ($\propto V^2$), and because this is independent of the charge, $u_2 = u_2({\cal G})$.

We now let $V=V_{bg} - V_{DC} + V_{AC}\sin{\omega t}$, where $V_{bg}$ is any background potential difference in the absence of charge added from CE (\textit{e.g.}~related to the contact potential differences or vertically separated bulk charges) and $V_{DC/AC}$ are the DC/AC driving voltages. We remark that the sign of $V_{DC}$ may change depending on the convention of a particular AFM. Taking the negative derivative of Eq.~\ref{eq:energy} with respect to the tip deflection, $z$, gives the vertical force, which can be separated into a DC component, a component at $\omega$, and a component at $2\omega$. Denoting $z$-derivatives as primed, the $\omega$ component is
\begin{equation}
    F_{\omega} = -u_1'qV_{AC}\sin{\omega t} - 2 u_2'(V_{bg}-V_{DC})V_{AC}\sin{\omega t}
    \label{eq:osc_force}
\end{equation}
In AM-KPFM, the quantity measured is the value of $V_{DC}$ that minimizes the oscillation amplitude, or equivalently nullifies the force, at $\omega$. Setting $F_{\omega}=0$ and solving for $V_{AM}\equiv V_{DC}$, we have
\begin{equation}
    V_{AM} = \frac{1}{2}\frac{u_1'q}{u_2'}+ V_{bg}.
    \label{eq:V_kpfm}
\end{equation}
Thus the presence of a point charge modifies the AM-KPFM voltage of a neutral insulator surface by the addition of the term $\tfrac{1}{2} u_1'q/u_2'$. 

We now extend Eq.~\ref{eq:V_kpfm} to account for a continuous surface charge density, $\sigma(\bld{r_q})$. Considering Eq.~\ref{eq:energy}, a term like $u_0 q^2$ will be present but now encapsulates all surface-charge parcels, $\sigma(\bld{r_q}) dx_q dy_q$, interacting with all their images. This term falls out of the analysis as it does not contribute anything to $F_\omega$. Second, the $u_2 V^2$ term is present and remains unchanged, since it does not depend on the surface charge. Third, the $u_1 q V$ term becomes the sum of the energies of each charge parcel interacting with the field of the capacitor, \textit{i.e.}~$u_1(|\bld{r_t}-\bld{r_q}|, {\cal G})qV \rightarrow \iint u_1(|\bld{r_t}-\bld{r_q}|, {\cal G}) \sigma(\bld{r_q})V dx_q dy_q$. Replacing $u_1'q$ in Eq.~\ref{eq:V_kpfm} with this integral form and defining $G_{AM}|\bld{r_t}-\bld{r_q}|, {\cal G}) \equiv \tfrac{1}{2} u_1'/u_2'$ yields
\begin{equation}
    V_{AM}(\bld{r_t})-V_{bg} = \iint \sigma(\bld{r_q})G_{AM}(|\bld{r_t}-\bld{r_q}|, {\cal G}) dx_q dy_q.
    \label{eq:kpfm_voltage}
\end{equation}
As this equation shows, the background-corrected voltage measured at tip position $\bld{r_t}$ is given by the convolution of the surface charge density with the appropriate Green's function governed by the system geometry.

The inverse problem can be solved by making use of the convolution theorem \cite{Gonzalez.2017}. Taking the Fourier transform of Eq.~\ref{eq:kpfm_voltage} and assuming the background is zero or corrected, we have
\begin{equation}
\hat{V}_{AM}\big{(} \bld{k} \, \big{)} = \hat{\sigma}\big{(} \bld{k} \, \big{)} \hat{G}_{AM}\big{(} \bld{k} \, \big{)}.
\label{eq:forward_convo}
\end{equation}
Solving for $ \hat{\sigma} \big{(} \bld{k} \, \big{)}$ and taking the inverse Fourier transform results in the charge density,
\begin{equation}
\sigma(\bld{r_q}) = {\cal F}^{-1} \bigg{\{}  \frac{\hat{V}_{AM}\big{(} \bld{k} \, \big{)}}{\hat{G}_{AM} \big{(} \bld{k} \, \big{)}}    \bigg{\}}.
\label{eq:backward_convo}
\end{equation}
\blue{As we mentioned earlier, a similar analysis can be done for FM-KPFM, which we explain in the Supplemental Material \cite{SupplMat}}.  In either case, the key to the problem is finding the Green's function. With the Green's function in hand, recovering the surface charge density is as straightforward as performing three Fourier transforms.

\section{\label{sec:green} Determining the Green's function with FEM simulations }

\begin{figure}[htb]
\centering
\includegraphics[width=0.45\textwidth]{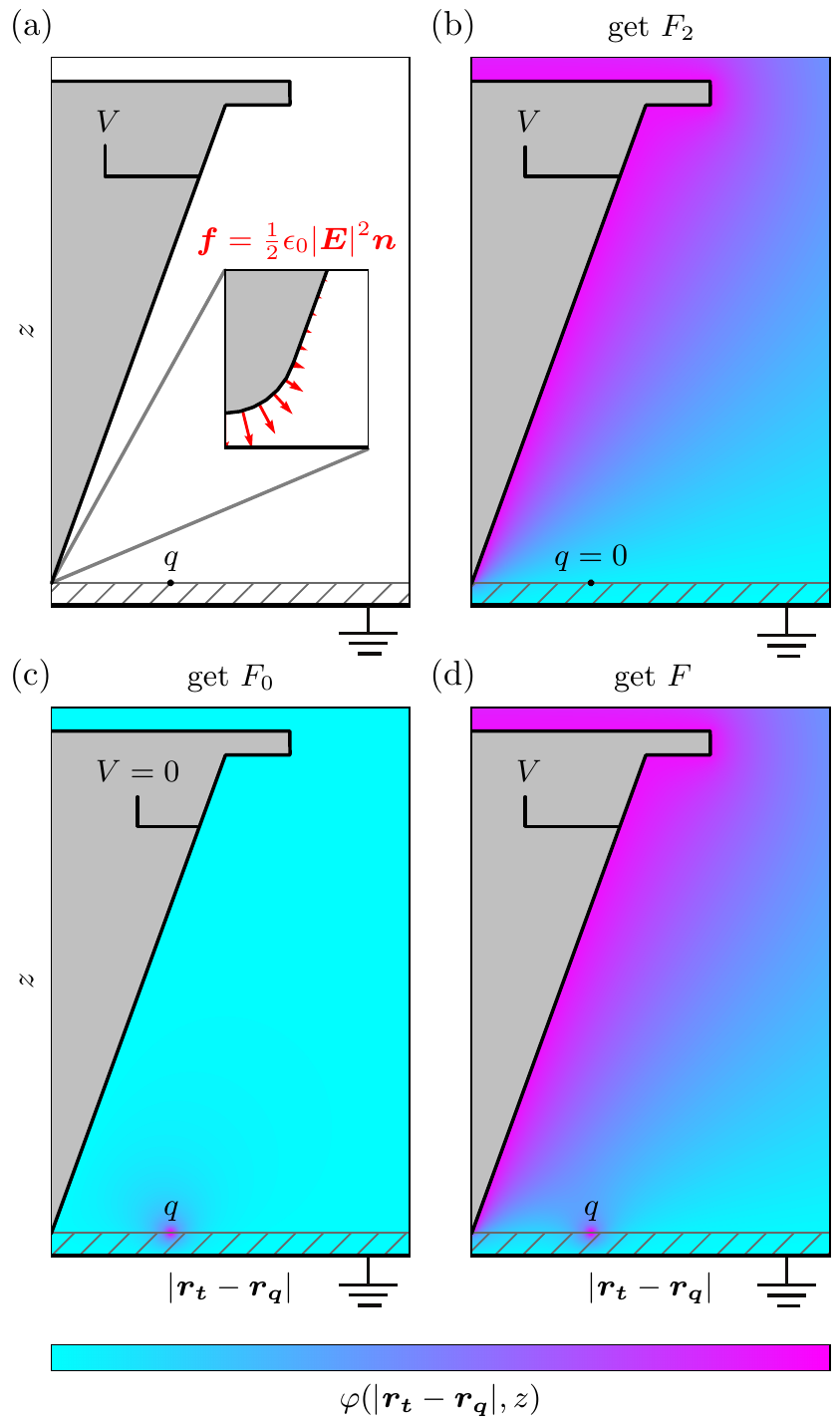}
\caption{ \label{fig:fem_greens_function} (a) We perform axisymmetric simulations in COMSOL to solve Poisson's equation with a `point' charge $q$ at $|\bld{r_t}-\bld{r_q}|$ and voltage $V$ at the probe/cantilever and 0 at the bottom plane. We approximate the cantilever as a disk of radius $R$. We integrate Maxwell's stress tensor over the probe/cantilever surface to calculate the force. For a particular $\{ |\bld{r_t}-\bld{r_q}|, {\cal G} \}$, we obtain: (b) $F_2$ in a simulation with $q=0$, (c) $F_0$ in a simulation with $V=0$, and then (d) $F$ in a simulation with both $q$ and $V$ nonzero to calculate $F_1=F-F_0-F_2$. We use $q$, $V$, $F_1$ and $F_2$ in Eq.~\ref{eq:greens_f_ratio} to determine the Green's function.
}
\end{figure}

Obtaining $G_{AM}$ amounts to knowing what the functions $u_1'(|\bld{r_t}-\bld{r_q}|, {\cal G})$ and $u_2'(|\bld{r_t}-\bld{r_q}|, {\cal G})$ are, but long-range electrostatics and geometric complexities make this exceptionally difficult. As previously mentioned, a common heuristic approach is to assume the charged surface and ground plane form a parallel plate capacitor, and that $V_{AM}$ is the (probe/cantilever free) voltage difference across this, yielding $\sigma\approx V_{AM} \kappa \epsilon_0 / \delta$ \cite{Barnes.2016, Wei:2016sr}. \blue{Although this relationship is intuitively appealing, we cannot find any reference that provides a rigorous derivation for it, and in the best case scenario it would only apply when the lateral extent of charged features is much larger than the thickness of the insulator, which for the heterogeneous charge features in nanoscale CE is almost never the case \cite{Baytekin.2011}.} More rigorously, Ref.~\cite{Orihuela.2016} made analytical headway by approximating the probe/cantilever with just the spherical tip and the insulator layer as infinitely thick. They then used the method of images to determine $u_1'$ and $u_2'$ from an infinite series of point charges, thus obtaining \textit{a} Green's function. \blue{However, this method is necessarily inaccurate because it ignores the vast majority of the AFM geometry \cite{Melin.2010, Guriyanova:2010}.}

We overcome geometrical complexity by obtaining the Green's function in FEM simulations, using COMSOL to solve Poisson's equation in the 2D axisymmetric geometry shown in Fig.~\ref{fig:fem_greens_function}(a). The features of the insulator layer, tip, and cone are the same as in Fig.~\ref{fig:physical_setup}. Instead of a continuous charge distribution, $\sigma(\bld{r_q})$, we consider a `point' charge at a distance relative to the tip, $|\bld{r_t}-\bld{r_q}|$. Due to the axis symmetry, our `point' is actually a ring of charge, but superposition renders the two equivalent. The one departure we make from Fig.~\ref{fig:physical_setup} is  approximate the (potentially tilted) cantilever as a disk of radius $R$. As we will show, this is justified because the cantilever almost exclusively affects the magnitude of the Green's function, but not spatial information. We limit the simulation volume to a radial distance $D>>R$ and use infinite element domains on the lateral/top boundaries.

\blue{As we will show momentarily, $G_{AM}$ can be extracted from our simulations by considering different forces in the system.  We calculate the vertical force on the simulated probe/cantilever by integrating the Maxwell stress tensor over the surface. Absent magnetic fields, the tensor reduces to $\mathbb{T} = \epsilon_0 (E_i E_j - \tfrac{1}{2}\delta_{ij}E^2)$. Furthermore, since the cantilever/probe are conductors, the field is always normal to their surfaces, hence the integral for the total vertical force reduces to $F = \bld{e_z} \cdot \oint \tfrac{1}{2}\epsilon_0 |\bld{E}|^2\bld{n} \, da$ [Fig.~\ref{fig:fem_greens_function}(a)]. Referring to Eq.~\ref{eq:energy}, we can decompose the total force as $F=F_0+F_1+F_2$, where $F_0 = -u_0'q^2$, $F_1=-u_1'qV$, and $F_2=-u_2'V^2$. Hence for a particular $q$ and $V$, the Green's function can be written in forces (instead of $u_1'$, $u_2'$) as
\begin{equation}
    G_{AM}(|\bld{r_t}-\bld{r_q}|, {\cal G}) = \frac{1}{2}\frac{V}{q}\frac{F_1}{F_2}.
    \label{eq:greens_f_ratio}
\end{equation}
We can isolate $F_1$ and $F_2$ as illustrated in Fig.~\ref{fig:fem_greens_function}(b-d). First we perform a simulation with $q=0$ and $V\ne0$ . In this case, the force on the probe is just $F_2$ [Fig.~\ref{fig:fem_greens_function}(b)]. To obtain $F_1$, we retain this value of $F_2$ and perform two more simulations. In one we set $V=0$ and $q\ne0$ to obtain $F_0$ [Fig.~\ref{fig:fem_greens_function}(c)]. In the other we use the same non-zero values previously used for $q$ and $V$ to get the full force, $F$. Using the calculated values for $F$, $F_0$, and $F_2$, we find $F_1=F-F_0-F_2$ [Fig.~\ref{fig:fem_greens_function}(d)]. Putting this all together, the AM-KPFM Green's function for a particular geometry at a particular probe/charge separation is then obtained via Eq.~\ref{eq:greens_f_ratio}. In this procedure, the exact non-zero values of $q$ and $V$ are not important. So long as we use the same values, the factor $V / q$ in front of Eq.~\ref{eq:greens_f_ratio} ensures that $G_{AM}$ is appropriately scaled. We point out that $G_{AM}$ is a negative function as we have defined it---$F_2$ is always negative (downward), and the factor $VF_1/q$ is always positive (upward). Combined with our AFM convention for the sign of $V_{DC}$, this means that positive (negative) surface charges produce negative (positive) voltages. Our procedure is similar to what Ref.~\cite{Melin.2010} used to do the forward problem of determining one single AM-KPFM voltage centered above a charged disk. By instead considering the effect of point charges to get the Green's function, we unlock the capacity to solve the inverse problem and determine the charge density from an arbitrary voltage map. }

\begin{figure}[t!]
\centering
\includegraphics[width=0.48\textwidth]{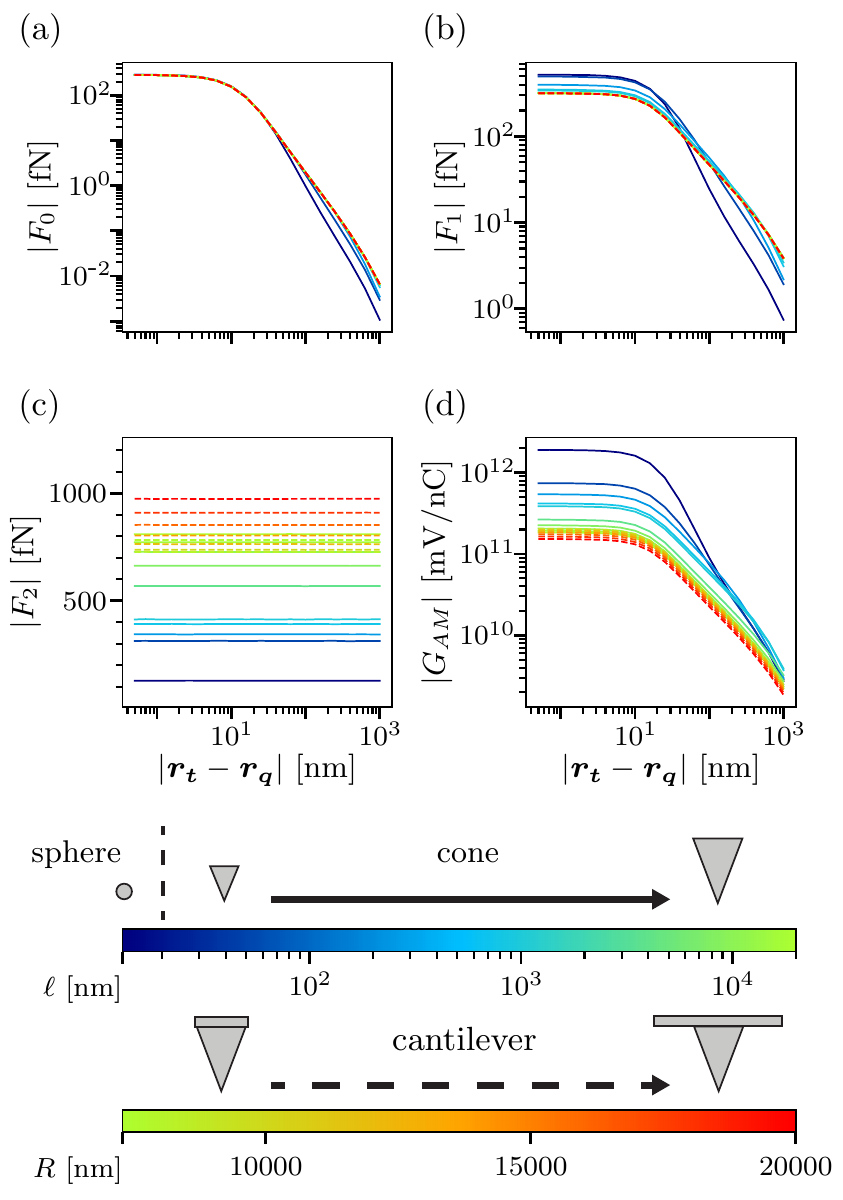}
\caption{ \label{fig:fem_gf_evolution} Evolution of (a) $|F_0|$, (b) $|F_1|$, (c) $|F_2|$, and (d) $|G_{AM}|$ starting from a spherical tip and ending with a full cone (solid curves), and then from a cone with small effective cantilever to a large one (dashed curves). The transition from sphere to cone affects both the magnitude and shape of all curves, whereas the cone-to-cantilever transition primarily affects the magnitude of $F_2$, and hence via Eq.~\ref{eq:greens_f_ratio} the magnitude of $G_{AM}$. Parameters common to all curves are $\delta = \SI{1}{\um}$, $a=\SI{20}{\nm}$, $d=\SI{10}{\nm}$, $\theta=\ang{20}$, $\kappa = 2.0$, and $T=\SI{1}{\um}$.
}
\end{figure}

In Fig.~\ref{fig:fem_gf_evolution} we plot $|F_{0,1,2}|$ and $|G_{AM}|$ as the geometry evolves from a spherical tip to a full cone, and then from a cone with a small effective cantilever to a large one. The sphere-to-cone evolution affects all terms. For the `charge' terms ($F_{0,1}$) it changes the magnitude and, in the earlier stages (up to $l\sim\SI{1}{\um}$), the shape of the curves. For the capacitive term ($F_2$) the shape is necessarily constant, but the magnitude significantly increases. During the cantilever evolution, the charge terms remain stable while the capacitive term continues to increase. To demonstrate how little the cantilever evolution affects $F_1$, we calculate the deviation between a full cone with no cantilever and one with $R=\SI{20}{\um}$, which is less than $0.5$\% averaged along the two curves.
The rationale for these observations is as follows. First, the $F_1$ term is large right under the tip because this is where the field of the capacitor is strongest. Similarly, $F_0$ is large here because this is where the real charge is closest to its images.  Both $F_0$ and $F_1$ decay rapidly moving away from the tip, with the steepest changes occurring at the radius, $a$. Second, $F_2$ depends only on the capacitive attraction between the probe/cantilever and ground, hence it continues to grow whenever more surface is considered. The takeaway is that spatial information in $G$ depends almost exclusively on the spherical tip and initial cone evolution (through $F_1$), whereas the magnitude information is ultimately modulated by the cantilever (through $F_2$). This justifies approximating a real cantilever with a disk, as it permits calibration to find an $R$ such that the Green's function has the correct magnitude, without adverse effects to spatial information \cite{SupplMat}. Strictly speaking, this convenience is only possible because $a,d<<l$, but this is true in most KPFM experiments.

\section{\label{sec:synthetic} Application of the method to synthetic data}

\begin{figure}[ht!]
\centering
\includegraphics[width=0.48\textwidth]{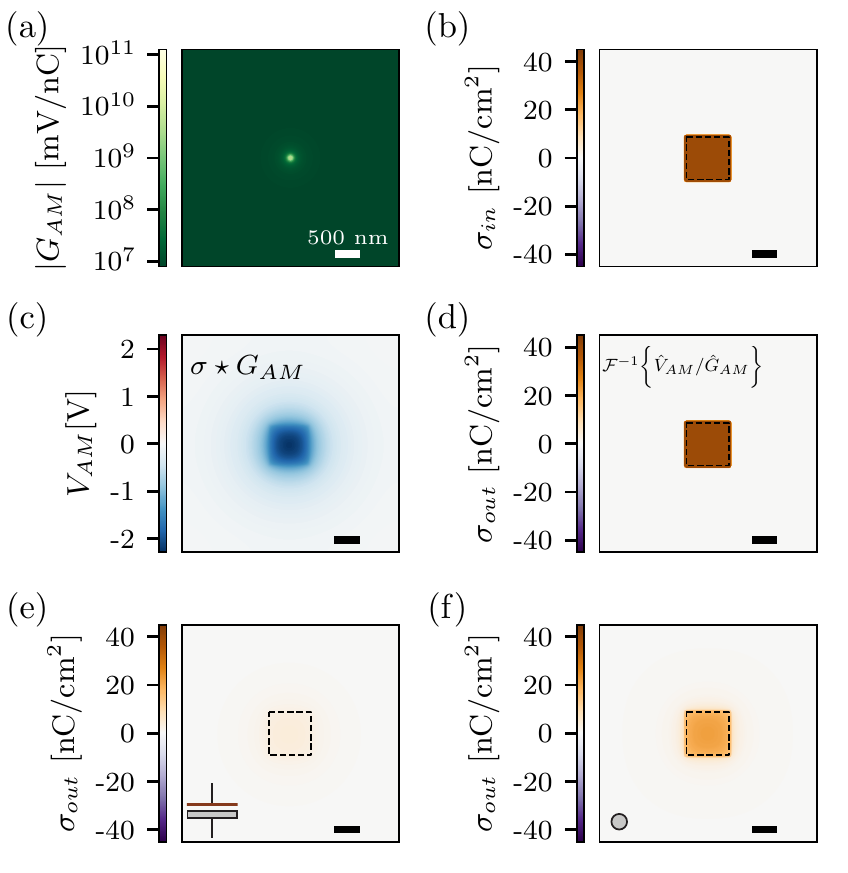}
\caption{ \label{fig:simulated_conversion} (a) Conversion of a numerically calculated, 1D Green's function [as in Fig.~\ref{fig:fem_gf_evolution}(d)] to a radially symmetric 2D map. Geometry is the same as in Fig.~\ref{fig:fem_gf_evolution} except with fixed $\ell=\SI{10}{\um}$ and $R=\SI{10}{\um}$. (b) Synthetic input charge map with side lengths $s=\SI{1}{\um}$ and charge density $\sigma=\SI{40}{\nano\coulomb/\cm^2}$. (c) Forward convolution of surface charge with Green's function to generate voltage map. (d) Deconvolution of surface charge with Green's function to exactly recover input. (e) If instead we use the `capacitor method' to recover the charge, it is grossly underestimated; color bar same as (d) to highlight this inaccuracy, with recovered charge so small ($>$ $\times$10) it is barely visible. Additionally, sharp spatial features of original charge distribution are lost. (f) If we deconvolve the voltage map with a Green's function corresponding to just the spherical tip, we do better with the spatial features, but still significantly miscalculate the original charge density.
}
\end{figure}
Before applying our method to experimental data, we perform tests with synthetic data. The advantage of this is we know what the input surface charge is, \blue{and can therefore compare the accuracy (\textit{i.e.}~errors relative to the true value) of the different charge recovery methods.} A preliminary step in either case is to convert our 1D Green's function into a 2D image, with dimensions set by the scan size and pixel length scale. We define the center as the origin and fill in pixel values according to Eq.~\ref{eq:greens_f_ratio}, assuming an $N\times N$ image, where $N$ is an odd number. Pixels whose $|\bld{r_t}-\bld{r_q}|$ falls between calculated points are filled with interpolation [Fig.~\ref{fig:simulated_conversion}(a)].
In Fig.~\ref{fig:simulated_conversion}(b), we show a synthetic charge density map consisting of a square with $\sigma=\SI{40}{\nano\coulomb/\cm^2}$ on a charge-free background. The side lengths of the square are $s=\SI{1}{\um}$, comparable to the size of experimentally observed contact-charge features \cite{Baytekin.2011}. The thickness of the insulator is also $\delta=\SI{1}{\um}$. We convert charge density to voltage via the forward convolution of $\sigma$ and $G_{AM}$ [Eq.~\ref{eq:forward_convo}, Fig.~\ref{fig:simulated_conversion}(c)]. We then use our method (\textit{i.e.}~deconvolving with the correct Green's function) to recover exactly the original input [Fig.~\ref{fig:simulated_conversion}(d)], \blue{\textit{i.e.}~the relative error of our method is $\sim$0\% }. By design, this test is tautological---it illustrates that our method works in the idealized case. We gain further insight by comparing this accuracy to what happens if the other, existing charge estimation methods are applied instead. If we use the `capacitor method' ($\sigma\sim V_{AM} \kappa \epsilon_0 / \delta$), we recover an average charge density in the square of $\SI{3.2}{\nano\coulomb/\cm^2}$, \blue{($\sim$91\% relative error) [Fig.~\ref{fig:simulated_conversion}(e)]}. This gross underestimation occurs because, even if there is a rigorous derivation for the capacitor method somewhere, it can only be argued for when $s>>\delta$. Since $s\sim\delta$, presumed contributions from (missing) charges at longer lengthscales lead to an incorrect reduction. Moreover, the recovered shape is smeared due to loss of spatial information---proper deconvolution allows one to resolve clearer charge features than can be seen in the voltage map. If instead of using the correct Green's function we use one generated for just a spherical tip [Fig.~\ref{fig:simulated_conversion}(f)], we do better recovering spatial information, but still introduce significant error. The charge density in the square is $\SI{18.3}{\nano\coulomb/\cm^2}$ \blue{($\sim$51\% relative error).} As we learned from Fig.~\ref{fig:fem_gf_evolution}, the problem in this case is that by neglecting the vast majority of the AFM geometry, we improperly calculate the terms $F_1$ and $F_2$.  \blue{Thus existing methods can significantly miscalculate the magnitude of the true charge density. As we show in the Supplemental Material, in certain circumstances the KPFM voltage (and by extension the capacitor method) can even get the \textit{sign} of charge incorrect \cite{SupplMat}.  This occurs when a small region of charge of one sign is surrounded by a larger region of the opposite sign, resulting in a voltage map of a single sign.  The takeaway is that quantitative and sometimes even qualitative information about the surface charge density is not reliably obtained without a rigorous approach.}

\section{\label{sec:experimental} Application to experimental data }

\begin{figure}[ht!]
\centering
\includegraphics[width=0.475\textwidth]{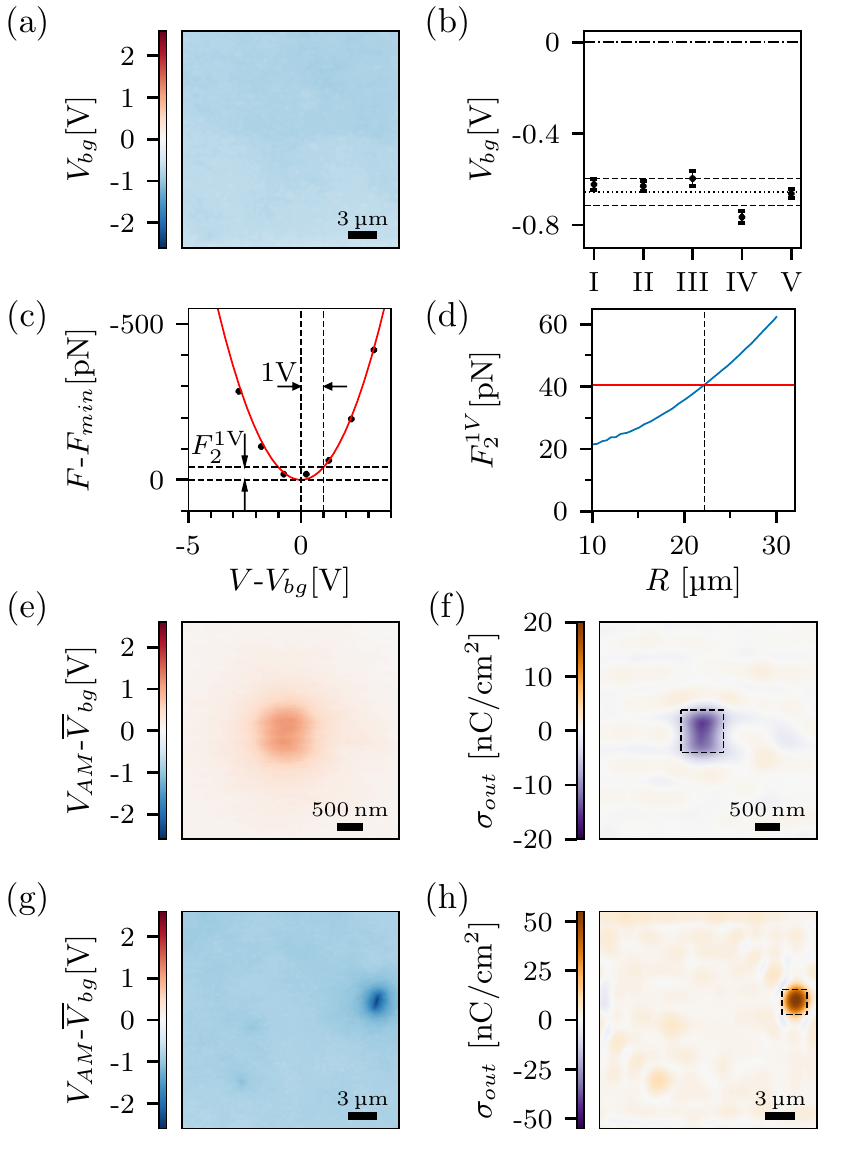}
\caption{ \label{fig:experiments} (a) AM-KPFM image of uncharged surface to obtain background $V_{bg}$. In this and subsequent voltage maps, we Fourier filter noise at length scales smaller than the tip radius, $a$. (b) Average $V_{bg}$ from 5 distant regions on the surface, illustrating it is uniform. Dotted lines represent global average and standard deviation. \blue{(c) Experimentally measured total force on cantilever \textit{vs.}~applied DC voltage. The surface is uncharged as in (a,b). We define and extract the force 1 V away from the minimum as $F_{2}^{1V}$. (d) We use $F_2^{1\text{V}}$ from (c) (red horizontal line) and compare it with the capacitive force for simulated probes with growing effective cantilever radii (blue line), also at 1 V.  We find the best $R$ where the two lines intersect (vertical dashed line).  For all panels in this figure, the calibrated cantilever disk has the same value, \blue{$R = \SI{22.2\pm0.4}{\um}$}}.  (e) Background subtracted voltage map of surface after charging via scraping a square of side length $\SI{1}{\um}$. (f) Recovered surface charge density from (e), where deconvolution reveals average value \blue{$\SI{-9.8\pm0.4}{\nano\coulomb/\cm^2}$} in the marked region. (g) Background corrected voltage map for different region of surface after 15 contacts with a macroscopic PDMS substrate. (h) Recovered surface charge from (g), showing a mean charge density of \blue{$\SI{2.2\pm0.2}{\nano\coulomb/\cm^2}$ and high charge regions with $\SI{25.2\pm0.6}{\nano\coulomb/\cm^2}$.}
}
\end{figure}

\blue{We now demonstrate application of our method, performing contact electrification experiments on a \ce{SiO_2} insulating layer with $\delta=\SI{3}{\um}$ and $\kappa\approx 4.2$. Our AFM is a Park Systems NX20, equipped with a MikroMasch NSC14/Cr-Au gold-coated probe with $\beta = \ang{13}$. This probe is pyramidal rather than conical, so we use the average cone half angle, $\theta\approx \ang{20}$. The cantilever parameters are $\ell=\SI{14}{\um}$, $a=\SI{55}{\nm}$ $L=\SI{125}{\um}$, $W=\SI{32}{\um}$ and $T=\SI{2.1}{\um}$, as measured with scanning electron microscopy. All experiments were performed in single-pass, AM mode with AC modulation frequency 17 kHz and offset heights of either $d=\SI{19}{\nm}$ [Fig.~\ref{fig:experiments}(a,b,g,h)] or $\SI{15}{\nm}$ [Fig.~\ref{fig:experiments}(e,f)]. The relative humidity is held constant during the experiments by means of controlled flow of dry nitrogen gas into the acoustic enclosure of the AFM, either at $36$ \% for Fig.~\ref{fig:experiments}(a,b,g,h) and $10$ \% for Fig.~\ref{fig:experiments}(e,f).} 

The first step is to find the background voltage, $V_{bg}$, that is present before the addition of $\sigma(\bld{r_q})$ during CE. We discharge samples by first placing them in an X-ray discharge chamber and then baking them at $\ang{200}$C for several hours. Subsequent measurements in a Faraday cup confirm this process leads to samples with zero net charge, and voltage maps at several locations confirm the surface is uniform [Fig.~\ref{fig:experiments}(a,b)]. \blue{These steps validate our assumptions about trapped bulk charges prior to CE---they are reduced to a small enough level to be negligible.} In the experiments described below, the background value used is from the exact region of interest when background/CE measurements at the same location were possible [Fig.~\ref{fig:experiments}(e,f)]. When this was not possible [Fig.~\ref{fig:experiments}(g,h)], we used the global average from the several regions of Fig.~\ref{fig:experiments}(b). We remark that this value \blue{($\overline{V}_{bg}= \SI{-0.66\pm0.06}{\volt}$)} is close to what is expected for the difference between the work functions of the backing silicon electrode and gold tip ($\sim$$0.56$ eV)\cite{haynes2016crc}.
\blue{The next step is to calibrate $R$ so the effective cantilever disk mimics the real cantilever. To do this, we first experimentally measure the total force on the real cantilever/probe as an applied DC voltage, $V$, is varied, as shown in Fig.~\ref{fig:experiments}(c). The parabolic shape is due to the capacitive force term, $F_2$, from which we extract the value at 1 V away from the minimum, defined as $F_2^{1\text{V}}$. Next, we perform charge-free simulations to determine the capacitive force of a probe with effective cantilevers of different radii, $R$, as plotted in Fig.~\ref{fig:experiments}(b). The potential difference in these simulations is set to be 1 V; hence, plotting $F_2^{1\text{V}}$ in the same graph as a horizontal line, the intersection gives the calibration value. As the capacitive force depends on the geometry, this calibration must be done whenever the geometry changes. In the case of the experiments of Fig.~\ref{fig:experiments}, the only geometric parameter we had to alter was $d$ (from $15$ nm to $19$ nm).  Given this change was small, we found the same value \blue{($R = \SI{22.2\pm0.4}{\um}$)} for all panels.} After this calibration, we construct the Green's function as previously explained.

Now we perform contact electrification experiments. First, we reproduce the square feature of Fig.~\ref{fig:simulated_conversion} by scraping charge into the surface with the tip connected to ground, applying a force of $\SI{300}{\nano\newton}$ while scanning over a region of $\SI{1}{\um}$ \blue{in contact-mode AFM (not KPFM mode) and with no tip bias}. This results in the background-corrected voltage of Fig.~\ref{fig:experiments}(e) where, like Fig.~\ref{fig:simulated_conversion}(f), we see a dense feature surrounded by a diffuse halo. Deconvolving this with our Green's function, we recover the surface charge density  \blue{$\sigma = \SI{-9.8\pm0.4}{\nano\coulomb/\cm^2}$}. We calculate the uncertainty with Monte-Carlo error propagation including contributions of $V_{bg}$ and $F_2^{1V}$. We obtain negative charge (positive voltage) on the surface, which is consistent with previous results of \ce{SiO_2} contacting metals \cite{lowell1990contact}. As shown in the Supplemental Material \cite{SupplMat}, applying the capacitor method, as it did for the synthetic data, severely underestimates the charge density, yielding \blue{$\SI{-1.1\pm0.1}{\nano\coulomb/\cm^2}$ (89\% relative deviation from our method)}, and using sphere-based Green's function yields \blue{$\SI{-6.6\pm0.1}{\nano\coulomb/\cm^2}$ (33 \% relative deviation).}

Next, we perform macroscopic charge transfer experiments with a $\SI{1}{\cm^2}$ PDMS counter sample (Sylgard\textsuperscript{TM} 184, 10:1 mixing ratio). We move the \ce{SiO_2} sample stage out from underneath the probe and perform a hand-pressed contact with the PDMS. We then return the sample stage and perform new AM-KPFM measurements. Fig.~\ref{fig:experiments}(g) shows an example background-corrected voltage map. Consistent with their expected places on the triboelectric series, the voltage for the \ce{SiO_2} is negative, indicating the presence of positive charge. In contrast to other results \cite{Baytekin.2011}, we see no features of alternating charge polarity---the surface is positively charged everywhere, though with heterogeneous `bright spots' of elevated intensity. Deconvolving this with the Green's function, we find that the average surface charge density is \blue{$\SI{2.2\pm0.2}{\nano\coulomb/\cm^2}$}, with a high-density feature of \blue{$\SI{25.2\pm0.6}{\nano\coulomb/\cm^2}$} at a length scale of $\sim\SI{2}{\um}$ [dashed square in Fig.~\ref{fig:experiments}(h)]. In this case, the average we obtain from capacitor method \blue{($\SI{1.1\pm0.1}{\nano\coulomb/\cm^2}$, 50 \% relative deviation)} is more consistent with the rigorous result, presumably since the length scale of the average charge transfer is much larger than the thickness of the \ce{SiO_2}. Importantly, however, the capacitor method still fails to recover the charge of features with small lateral scale---for the high density region, as expected it again yields a smaller value \blue{$\SI{2.1\pm0.1}{\nano\coulomb/\cm^2}$ (92 \% relative deviation)}. 

\section{\label{sec:conclusion} Conclusions}

\blue{We have introduced a rigorous method to extract surface charge density from KPFM voltage maps. 
Conceptually, our work reiterates \cite{Orihuela.2016, Gonzalez.2017} that the key to the problem is to find the appropriate Green's function, which makes recovering charge a simple matter of deconvolution. Practically, we overcome the geometric complexity involved in calculating this Green's function by relating it to forces that are obtainable in FEM simulations.
The entire process takes approximately 10 minutes on a contemporary computer. Although we have focused on AM-KPFM, we show in the Supplemental Material, a similar approach is possible for FM-KPFM \cite{SupplMat}.  Our main approximation is to replace the rectangular cantilever as a disk in our FEM simulations. As we have shown, this is justified because the effect of the cantilever is almost exclusively to change the magnitude of the Green's function, but not spatial information---hence by calibration an appropriate disk radius can be determined to yield the correct magnitude. We have shown that existing methods can grossly miscalculate the magnitude of the charge density, and in certain instances the KPFM voltage itself can even misrepresent the correct sign of charge. Our experiments illustrate the capacity of our method quantitatively extract charge and to see finer features than with the voltage map alone. Our method assumes rotational/translational symmetry in the geometry, and that the charge to be measured is close to the surface compared to the tip/sample distance, but these are among the most common situations. With sufficient computational power to efficiently calculate symmetry-reduced Green's functions for different locations, these limitations could be overcome to address more complex situations.}

\begin{acknowledgments}
This project has received funding from the European Research Council (ERC) under the European Union’s Horizon 2020 research and innovation programme (Grant agreement No.~949120). This research was supported by the Scientific Service Units of The Institute of Science and Technology Austria (ISTA) through resources provided by the Miba Machine Shop, the Nanofabrication Facility, and the Scientific Computing Facility. We thank Florian Stumpf from Park Systems for useful discussions and support with scanning probe microscopy.

F.~Pertl and J.C.~Sobarzo contributed equally to this work.
\end{acknowledgments}

\bibliographystyle{apsrev4-2}
\bibliography{manuscript}% Produces the bibliography via BibTeX.

\end{document}

\setcounter{figure}{0}
\renewcommand{\figurename}{SUPPL.~FIG.}
\renewcommand\thefigure{\arabic{figure}}
\setcounter{table}{0}
\renewcommand{\tablename}{SUPPL.~TABLE}
\renewcommand\thetable{\arabic{table}}

\begin{center}
{\huge{ \textbf Supplemental Information}}
\end{center}

 \section{Extension of the method to FM-KPFM}
 
\blue{We use AM-KPFM as opposed to FM-KPFM because it works better for our thick ($\SI{3}{\um}$) substrates.  However, our method can be adapted for FM-KPFM. In FM-KPFM, the voltage measured is the value of $V_{DC}$ that sets $F_{\omega}'=0$.  Considering a single charge on the surface (Eq.~1), $F_{\omega}'$ is given by
\begin{equation}
    F_{\omega}' = -u_1''qV_{AC}\sin{\omega t} - 2  u_2''(V_{bg}-V_{DC})V_{AC}\sin{\omega t}.
    \label{eq:fore_gradient}
\end{equation}
Setting this equal to zero and solving for $V_{FM}\equiv V_{DC}$ yields the FM-KPFM voltage for the single charge
\begin{equation}
    V_{FM} = \frac{1}{2}\frac{u_1''q}{u_2''}+ V_{bg}.
    \label{eq:V_kpfm}
\end{equation}
By the same reasoning we laid out for the AM case, we conclude that the Green's function for FM-KPFM case is therefore $G_{FM}(|\bld{r_t}-\bld{r_q}|, {\cal G}) \equiv \tfrac{1}{2} u_1''/u_2''$. Analogously to Eq.~7 of the main text, we can rewrite this in a way that is amenable to parameters that can be calculated in FEM
\begin{equation}
    G_{FM} = \frac{1}{2}\frac{V}{q}\frac{F'_1}{F'_2}.
    \label{eq:greens_f_ratio}
\end{equation}
This could be implemented by simulating the forces $F_{0,1,2}$ at one vertical position, and then again for a small displacement $\delta z << \delta, d, a$. The calibration for the effective cantilever would be the same in AM mode as in FM mode.
Hence, obtaining $G_{FM}$ in FEM simulations is not as quite as straightforward as in the AM-KPFM case (requiring twice the number of simulations), but still possible. Finally, once one has the Green's function then backing out charge from the voltage map is accomplished with the Fourier transforms of Eq.~6 and the appropriate $G_{AM/FM}$, depending on the case.}
 
 \section{Additional plots relevant to the main text}

In Suppl.~Fig.~\ref{fig:suppl_fig2}, we show the charge densities recovered from our experimental data for the capacitor method and a sphere-based Green's function.
Panels (a,b) show the results for the square of charge scratched by the AFM tip. While the average charge density in the square recovered from the proper Green's function is \blue{$\sigma = \SI{-9.8\pm0.4}{\nano\coulomb/\cm^2}$}, the values recovered for the capacitor/sphere methods are \blue{$\SI{-1.1\pm0.1}{\nano\coulomb/\cm^2}$ and $\SI{-6.6\pm0.1}{\nano\coulomb/\cm^2}$}, respectively.  The deviation of the capacitor method in this case is even worse, presumably due to the fact that the square side length ($\SI{1}{\um}$) is now significantly less than the insulator thickness ($\SI{3}{\um}$). 
\begin{figure}[!ht]
\centering
\includegraphics[width=0.5\textwidth]{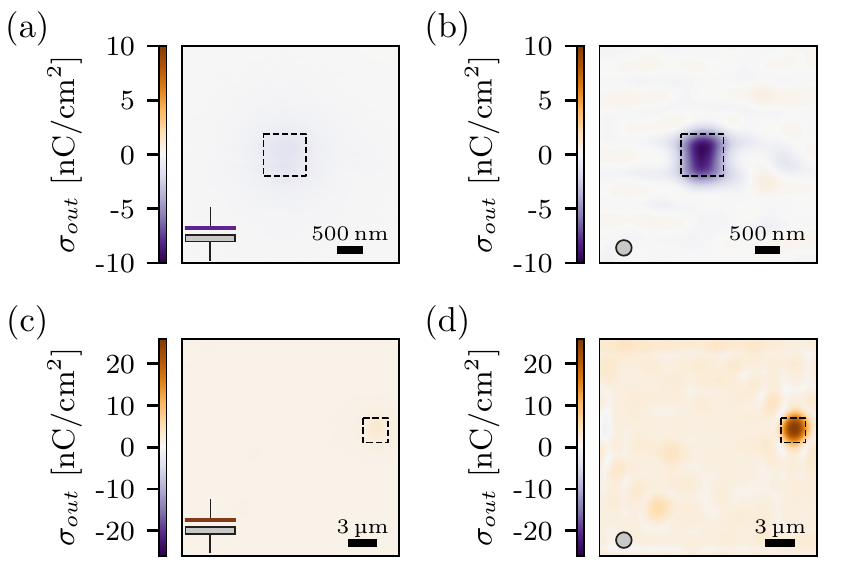}
\caption{\label{fig:suppl_fig1} (a) `Capacitor' recovered surface charge density from Fig.~5(c) with average of \blue{$\sigma = \SI{-1.1\pm0.1}{\nano\coulomb/\cm^2}$} in the dashed region. (b) `Sphere' recovered surface charge density of the same with $a=\SI{55}{nm}$ and $d=\SI{19}{nm}$, where deconvolution reveals average of \blue{$\sigma = \SI{-6.6\pm0.1}{\nano\coulomb/\cm^2}$}. (c) Capacitor recovered surface charge density from Fig.~5(e) with global average \blue{$\SI{1.1\pm0.1}{\nano\coulomb/\cm^2}$} and bright-spot average \blue{$\SI{2.1\pm0.1}{\nano\coulomb/\cm^2}$} in the dashed region. (d) Sphere recovered surface charge  density ($a=\SI{55}{nm}$ and $d=\SI{19}{nm}$) with global average \blue{$ \SI{2.0\pm0.1}{\nano\coulomb/\cm^2}$ and dashed-region average $ \SI{13.2\pm0.1}{\nano\coulomb/\cm^2}$.}}
\end{figure}

Panels (c,d) show the results for the contact against PDMS. Here we investigate two separate aspects of the charge recovery.  First, the background charge density calculated with the proper Green's function [light red shade of Fig.~5(f)] is \blue{$\sigma = \SI{2.2\pm0.2}{\nano\coulomb/\cm^2}$}.  Due to the fact that the lateral extent of the background is $>>\delta$, the capacitor approximation is more likely to be applicable, and hence from it we recover \blue{$\SI{1.1\pm0.1}{\nano\coulomb/\cm^2}$}. However, in the bright red spot, whose lateral extent is small ($\SI{2}{\um}$), the capacitor method again underestimates (\blue{$\SI{2.1\pm0.1}{\nano\coulomb/\cm^2}$ \textit{vs.}~$ \SI{25.2\pm0.6}{\nano\coulomb/\cm^2}$ when calculated properly}).  For completeness, the recovered values of the background and bright spot in the sphere recovery are \blue{$\SI{2.0\pm0.1}{\nano\coulomb/\cm^2}$ and $\SI{13.2\pm0.1}{\nano\coulomb/\cm^2}$, respectively.}

\section{How KPFM voltages can misrepresent the sign of charge}

\begin{figure}[h!]
\centering
\includegraphics[width=0.5\textwidth]{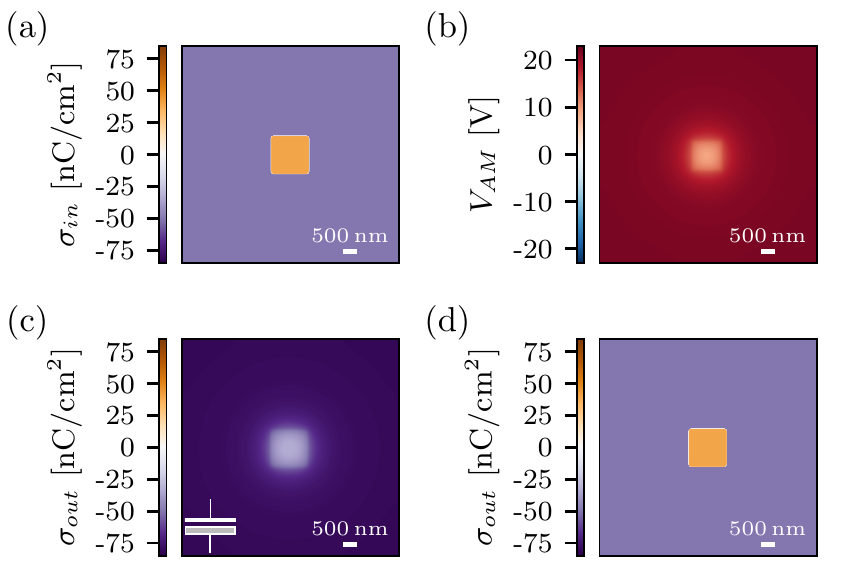}
\caption{ \label{fig:suppl_fig2} \blue{ (a) Synthetic input charge map with a side length of $s=\SI{1}{\um}$ and charge density of $\sigma=\SI{40}{\nano\coulomb/\cm^2}$ and surrounding charge density of $\sigma=\SI{-50}{\nano\coulomb/\cm^2}$. (b) Forward convolution of surface charge with Green's function to generate voltage map similar to Fig.~4(c). (c) Deconvolution of surface charge with the 'capacitor method' leads to miscalculation of charge sign. (d) If instead we use the proper Green's function to recover the charge, we obtain the identical charge map as in (a).}}
\end{figure}

\blue{As mentioned in the main text, there are certain circumstances where the KPFM voltage cannot be trusted to give the correct sign of charge. This occurs when a small charge feature of one sign is surrounded by a much larger charge feature of the opposite sign. For example, in Suppl. ~Fig.~\ref{fig:suppl_fig2}(a) we show a positively charged feature of side length of $s=\SI{1}{\um}$ and $\sigma=\SI{40}{\nano\coulomb/\cm^2}$ surrounded by a much larger negatively charged region of $\sigma=\SI{-50}{\nano\coulomb/\cm^2}$. If we convolve this charge map with a realistic Green's function we obtain the voltage map in Suppl.~Fig.~\ref{fig:suppl_fig2}(b). We see that the positive charge in the center is screened by the negative surrounding charge, \textit{i.e.}~it has a voltage above it of the \textit{same sign} as the surrounding region. If we use the capacitor method to extract charge from this, we obtain not just the wrong magnitude, but also the wrong sign: $\sigma=\SI{-42.5}{\nano\coulomb/\cm^2}$ [Suppl.~Fig.~\ref{fig:suppl_fig2}(c)]. If instead we properly deconvolve the voltage map with the Green's function, we recover a high fidelity copy of the original input with the correct magnitude and sign [Suppl.~Fig.~\ref{fig:suppl_fig2}(d)]. This again makes the point that the capacitor method, even if it has a rigorous derivation somewhere, can only be expected to work for large charge features with $s >> \delta$, and that the sign extracted from KPFM voltage maps or the capacitor method cannot necessarily be trusted. }
